\documentclass[pra,showpacs,preprintnumbers,amsmath,amssymb,tightenlines,twocolumn,superscriptaddress]{revtex4}

\usepackage{graphicx}
\usepackage{dcolumn}
\usepackage{bm}
\usepackage{epsfig}
\usepackage{stmaryrd}
\usepackage{color}

\newcommand{\bra}[1]{\langle\,{#1}\, |}
\newcommand{\ket}[1]{|\,{#1}\,\rangle}

\newcommand{\vek}[1]{\boldsymbol{#1}}
\newcommand{\eref}[1]{Eq.~(\ref{#1})}

\newcommand{\cref}[1]{chapter~\ref{#1}}

\newcommand{\Cref}[1]{Chapter~\ref{#1}}

\usepackage{ulem}  
\begin{document}

\title{ The generalized imaging theorem: quantum to classical transition without decoherence} 
\author{John S.\ Briggs}
\affiliation{Institute of Physics, University of Freiburg, Freiburg, Germany}
\email{briggs@physik.uni-freiburg.de}
\author{James M.\ Feagin}
\affiliation{Department of Physics, California State University-Fullerton, Fullerton, CA 92834, USA  }
\email{jfeagin@fullerton.edu}

\begin{abstract}
The mechanism of the transition of a dynamical system from quantum to classical mechanics is one of the remaining challenges of quantum theory. Currently, it is considered to occur via decoherence caused by entanglement  and/or stochastic interaction with a quantum environment. Here we prove that, in the absence of de-phasing environmental interaction, the asymptotic spatial wave function of any quantum system, propagating over distances and times large on an atomic scale but still microscopic, and subject to arbitrary deterministic fields, becomes proportional to the initial momentum wave function,  \emph{where the particle positions and momenta are related by classical mechanics.}
This implies that detection of particle positions and momenta at different times will yield results in accordance with classical mechanics, without the need to postulate decoherence of the wave function.
\end{abstract}
\maketitle

It has become widely accepted that the reason we observe a classical deterministic world, although the motion of particles is governed by a probabilistic quantum description, can be attributed to the phenomenon of \emph{decoherence} \cite{decohere}. This is the change in the wave function of a quantum system due to interaction with a quantum environment, variously taken to be the ambient surroundings, a measuring apparatus, or a combination of both. One change necessary to achieve a classical status is shown to be the transition of a particle or particles from an entangled and delocalized state to an incoherent superposition of probabilities, rather than a coherent superposition of amplitudes. To represent a classically localized particle, the spatial wave function must become  extremely narrow so that the probability to detect the particle becomes correspondingly localized.  Even though a particle may travel macroscopic distances before detection, it is necessary that the localization due to decohering interaction overcomes the natural spreading of the wave function. It is assumed that spatial localization of a wave function is a prerequisite for classical behavior, and this is achieved by environmental decoherence. The deterministic propagation of the quantum system under the sole influence of its own Hamiltonian is considered not to lead to classical behaviour. Here we 
demonstrate that clear aspects of classical motion do appear from such propagation, including that due to applied external fields, typically over times and distances which are microscopic.  In particular, it is shown that the locus of points of equal quantum probability defines a classical trajectory. Hence classical motion will be inferred from particle detection at macroscopic distances.

The changes of a quantum system due to environment interaction are usually discussed in terms of the density matrix rather than the wave function. Typically \cite{decohere}, the equation for the time and space propagation of the density matrix is split into three contributions, namely, the ``von Neumann"  term due to propagation by the system Hamiltonian itself  plus two phenomenological terms arising from interaction with an environment. The first is a dissipative term, whose influence on the quantum to classical transition is usually considered negligible, describing loss of probability and/or energy into environment degrees of freedom. The important contribution is a stochastic, temperature-dependent interaction leading to decoherence. This is considered the driving term of the emergence of classical behavior. The loss of quantum coherence is signified by the off-diagonal elements of the density matrix
becoming zero. For classical particle attributes to appear the remaining diagonal elements must become localized in space. 

Without doubt the decoherence scenario describes correctly changes in the system wave function leading to quantum correlation (entanglement) being transformed into correlation manifest by classical particles. Also it is usually recognized that the deterministic von Neumann propagation of the density matrix, or equivalently the Schr\"odinger propagation of the wave function, also obeys Ehrenfest's theorem, which shows that the expectation value of any observable changes in time according to classical mechanics. However, for extensively de-localized wave functions, this is practically of little meaning and in no way implies that a quantum to classical transformation has been made. Ehrenfest's theorem implies classical characteristics only following localization due to decoherence.
Here we show that  aspects of the quantum to classical transition do arise from the undisturbed Schr\"odinger propagation of the system wave function without coupling to a quantum environment.  After propagation to distances which are still on the nanoscale, it is shown that  classical motion is  encoded in the wave function itself in that  \emph{each and every point} of the wave function at different times are connected by a classical relation between coordinates and momenta. This is a far more powerful statement than Ehrenfest's theorem involving  only averages.

The essential ingredient of our proof is simply to unify two well-known aspects of quantum dynamics whose connection hitherto has not been recognized. The first is that classical motion arises simply as a result of deterministic Schr\"odinger  propagation in that, for large asymptotic times and distances, the quantum propagator can be approximated by its semiclassical form. This form is decided purely by classical mechanics through the classical action function. Although the techniques are well developed and basic to the whole field of semiclassics and quantum chaos, the use of the semiclassical approximation has been confined mostly  to the approximate calculation of atomic and molecular bound and scattering \emph{quantum} states \cite{semicl, Brumer}.

The second aspect is a result from scattering theory, known as the \emph{imaging theorem} (IT). 
The IT has a long history, beginning with the work of Kemble in 1937  \cite{Kemble} who wished to identify the momentum of a freely-moving  collision fragment arriving at a detector. It has been derived for free motion several times since then, in connection with multi-particle quantum collision theory \cite{ITfree}.  Recently we have extended its validity to the extraction of collision fragments by constant electric and magnetic fields \cite{BriggsFeagin_IT}. The IT relates the coordinate space wave function at large distances from a collision region to the momentum space wave function at the boundary of the collision region and involves a  classical relation between position and momentum variables. Again, however, the precise connection to the semiclassical propagator has not been recognized.

Here we combine these two aspects in that we use the results of semiclassical quantum theory to generalize the IT to (nonrelativistic) motion under the influence of arbitrary external laboratory fields and particle interactions. Our generalized IT shows that the spatial wave function of any quantum system propagating over macroscopic distances and times becomes proportional to the initial momentum wave function,  \emph{where the position and momentum are related by classical mechanics}. Most importantly, this implies that the probability to measure a particle at a given position at a certain time is identically equal to the probability that it started with a given momentum at an earlier time and has moved according to a classical trajectory.  However, the full delocalized quantum wave function remains intact. Hence, a detection will infer classical behavior of the particle  \emph{without any environmental interaction whatsoever,} which would lead to decoherence. In short, an observer would conclude that the motion is classical despite it being governed by the Schr\"odinger equation. There is no change in the wave function apart from that due to external forces deciding the particle's quantum motion. In particular, the spatial wave function remains delocalized, or even entangled, and suitably designed measurements can still detect  quantum interference.

Of course the phenomenon of decoherence is connected with more philosophical arguments as to the meaning of the wave function, whether a particle is a wave, does the wave function ``collapse" during measurement, etc. Here we adopt a simple and pragmatic philosophy. A particle is always a particle. The wave function is simply an ``information field" whose squared modulus of the amplitude gives the probability to detect a particle at a certain position or momentum (Born's rule).  
From the IT, the results of detection of different particles at different phase space points will be compatible
with their classical motion, even though describable by a quantum wave function. Interestingly, this theoretical result is pre-supposed in countless experiments \cite{Helm2014} in which classical mechanics is used to calculate the motion of microscopic particles from an interaction volume of atomic dimensions to a detector of macroscopic dimensions placed at a macroscopic distance from the reaction volume.

To be precise, we consider a system of $n$ quantum particles described by $3n$ dimensional position
vector $\vek x$ or momentum vector $\vek p$. 
Quantum particles are to be understood as material particles whose size and energy are sufficiently small that their motion must be described by quantum mechanics.
The particles interact in a volume of microscopic dimensions and emanate, at time $t = t_i$, from this volume of interaction with a momentum distribution described by the wave function $\tilde\Psi(\vek p, t_i)$.
There follows propagation, usually under the influence of external forces and possible lensing systems, to a time $t = t_f$ and a point of detection  $\vek x(t_f) = \vek x_f$ at macroscopic distances from the reaction volume. 
The corresponding state of the system $\ket{\Psi(t_i)}$ propagates in time according to
$\ket{\Psi(t)} = U(t,t_i) \,\ket{\Psi(t_i)}$,
where $U(t,t_i)$ is the time-development operator. 
Projecting onto an eigenstate $\bra{\vek x_f}$ of final position $\vek x_f$ and inserting a complete set of momentum eigenstates, this propagation is expressed in terms of wave functions as
\begin{equation}
\Psi(\vek x_f,t_f)  = \int  d\vek p \, \tilde K(\vek x_f,t_f; \vek p,t_i) \,  \tilde\Psi(\vek p, t_i),
\label{Psi_mixed}
\end{equation}
where $\tilde K(\vek x_f,t; \vek p,t_i) = \bra{\vek x_f}U(t,t_i)\ket{\vek p}$ is the \emph{mixed} coordinate-momentum propagator. 
  
In principle, the propagator is described exactly by a Feynman path integral involving the action $ \tilde S(\vek x,t; \vek p,t_i)$.
Instead, let us assume that propagation has proceeded to phase space points 
$\vek x, \vek p,t$ where the action $\tilde S$ is large
compared to $\hslash$. 
For larger times we may approximate the propagator by the corresponding  semiclassical propagator in which the \emph{classical} action $ \tilde S_c(\vek x,t; \vek p,t_i)$ appears. The boundary of this transition zone from quantum to classical action is designated by  classically conjugate variables $\vek x_i, \vek p_i,t_i$.

The semiclassical mixed propagator is given by \cite{Brumer}
\begin{eqnarray}
\label{SemiK}
 \tilde K_{sc}(\vek x_f,t_f; \vek p,t_i) &=& (2\pi i \hslash)^{-3n/2} 
 	\left|\det \frac{\partial  \vek x_f}{\partial \vek x} \right|^{-1/2}  \nonumber \\
	&\times& \exp\left(\frac{i}{\hslash} \tilde S_c(\vek x_f,t_f; \vek p,t_i)  \right),
\end{eqnarray}
where we consider the contribution of only a single classical trajectory fully resolved by its time of flight.  Also, since later we concentrate on probabilities, we suppress a Maslov phase. 

We relate the mixed action $\tilde S_c(\vek x_f,t_f; \vek p,t_i)$ to the action in coordinate space
$ S_c(\vek x_f, t_f; \vek x ,t_i)$ by the  Legendre transformation
\begin{equation}
\tilde S_c(\vek x_f,t_f; \vek p,t_i) = S_c(\vek x_f, t_f;\vek x, t_i)  + \vek  p \cdot (\vek x - \vek x_i),
\end{equation}
where here $\vek x$ is considered a function of $\vek x_f$ and $\vek p$ and the times $t_f, t_i$. When the propagator \eref{SemiK} is substituted in Eq.\ (\ref{Psi_mixed}),
 the stationary phase of the integral is defined by 
 $\partial \tilde S_c/\partial \vek p = \vek x  - \vek x_i \equiv 0$,
and the root of this equation defines a stationary phase point $\vek p \to \vek p_i$.

Evaluating the integral in the stationary-phase approximation gives \cite{semicl}
\begin{eqnarray}
\label{PsiSPA}
\Psi(\vek x_f,t_f)  &\approx& (2\pi \hslash)^{3n/2} \, \left|\det \frac{\partial^2  \tilde S_c}{\partial \vek p_i \partial \vek p_i} \right|^{-1/2} \nonumber \\ 
&\times& \tilde K_{sc}(\vek x_f,t_f; \vek p_i,t_i) \,  \tilde\Psi(\vek p_i, t_i).
\end{eqnarray}
Here the determinant of the Hessian $\partial_{\vek p_i, \vek p_i }^2  \tilde S_c$ combines with the determinant in Eq.\ (\ref{SemiK}) to give the familiar 
Van Vleck determinant of the Jacobian $\partial^2_{\vek x_f, \vek x_i} S_c$ according to
\begin{equation}
\left|\det \frac{\partial  \vek x_f}{\partial \vek x_i} \right|^{-1/2} \left| \,
\det \frac{\partial^2  \tilde S_c}{\partial \vek p_i \partial \vek p_i} \right|^{-1/2} 
= \left|\det \frac{\partial^2 S_c}{\partial \vek x_f \partial \vek x_i} \right|^{1/2}. 
\label{VVdet}
\end{equation}
Thus we obtain \emph{for a single trajectory} of an $n$-particle system the asymptotic wave function
\begin{eqnarray}
\Psi(\vek x_f,t_f)  &  \approx & i^{-3n/2}
 	 \left|\det \frac{\partial^2 S_c}{\partial \vek x_f \partial \vek x_i} \right|^{1/2} \nonumber \\
	 &\times& e^{i S_c(\vek x_f, t_f; \, \vek x_i, t_i)/\hslash} \, \tilde\Psi(\vek p_i, t_i),
\label{ITwave}
\end{eqnarray}
where the coordinates $\vek x_f$ and $\vek x_i$ are connected by the classical trajectory defined by $ \vek p_i$. 

With this result we obtain the quantum generalization of the classical trajectory density $d \vek p_i/d \vek x_f$ of finding the system in the volume element $d\vek x_f$ given that it started with a momentum $\vek p_i $ in the volume element $d \vek p_i$ (see Gutzwiller \cite{semicl}, chap.\ 1). This is just given by the Van Vleck determinant, so that from Eq.\ (\ref{ITwave}) we have
\begin{eqnarray}
\frac{d \vek p_i}{d \vek x_f} =  \left|\det \frac{\partial  S_c}{\partial \vek x_f \partial \vek x_i} \right|  \, \approx \,  \frac{|\Psi(\vek x_f,t_f)|^2}{|\tilde\Psi(\vek p_i, t_i)|^2},
\label{VVtheorem}
\end{eqnarray}
where the wave function coordinates and momenta are connected by the plexus of classical trajectories from $d \vek p_i$ to $d\vek x_f$.
Thus we also connect the \emph{classical} propagation with the asymptotic quantum measurement probabilities,
\begin{eqnarray}
 |\Psi(\vek x_f,t_f)|^2 \,d \vek x_f  \, \approx \,   |\tilde\Psi(\vek p_i, t_i)|^2 \, d \vek p_i.
\label{ITprob}
\end{eqnarray}
Eqs.\ (\ref{ITwave}), (\ref{VVtheorem}) and (\ref{ITprob}) embody the generalized IT and are the main results of this paper \cite{k_measure}.

The extent of the quantum to classical transition described by the IT is seen by direct comparison of \eref{ITwave} and the exact \eref{Psi_mixed}. In the latter the relation between the noncommuting variables $\vek x$ and $\vek p$ is nondeterministic in that the spatial wave function at position $\vek x_f$ is given by a transform of the momentum wave function involving integration over all possible values of  $\vek p$. By contrast, in the IT of \eref{ITwave}, the spatial wave function at $\vek x_f$ is directly proportional to the momentum wave function at $\vek p_i$, where $\vek x_f$ and $\vek p_i$ are classical variables connected deterministically by the classical trajectory.
This connection can, in principle, be continued all the way in to the edge of the transition zone.
Hence it remains to consider how large is the limit of the zone beyond which classical motion is manifest and the IT is valid.

Let us consider the absolutely simplest case, that of a single particle of mass $m$ undergoing free motion in one dimension. Since all $x_i$ are of microscopic size and the $x_f$ are considered macroscopic, it suffices if one takes for simplicity, $x_i = 0, t_i=0$. Further defining $p_f = p_i \equiv p$, and $t_f \equiv t$, one has $x_f = p t/m$ and $dp_i/dx_f = m/t$. The IT then appears in the standard form \cite{ITfree,BriggsFeagin_IT}. 
\begin{equation}
\label{onedIT}
\Psi( x_f,t) = \left(\frac{m}{i t}\right)^{1/2} \exp\left[\frac{i}{\hslash}\frac{p^2}{2m}t\right] \tilde \Psi(p).
\end{equation} 
Consider initially a Gaussian of width $\sigma$  launched at time $t=0$ and described by
\begin{eqnarray}
\label{2wfns}
\Psi(x,0) &=& (\pi\sigma^2)^{-1/4} e^{-x^2/(2\sigma^2)}, \nonumber \\
\tilde\Psi(p) &=&  \left(\frac{\sigma^2}{\pi \hslash^2}\right)^{1/4} e^{-p^2 \sigma^2/(2\hslash^2)}.
\end{eqnarray}
For $t > 0$ the initial spatial wave function propagates freely in time and has the exact form
\begin{equation}
\begin{split}
\Psi(x,t) = & \left(\frac{\sigma^2}{\pi}\right)^{1/4} \left(\sigma^2 + \frac{i\hslash t}{m}\right)^{-1/2}\\
&\times \exp{\left[-\frac{x^2}{2} \frac{\sigma^2- i\hslash t/m}{\sigma^4+\hslash^2 t^2/m^2}\right]}.
\end{split}
\end{equation}  
The IT condition emerges in the limit of large times which here corresponds to $\hslash t/m \gg \sigma^2$. Then the spatial wave function assumes the form
\begin{equation}
\label{Psifree}
\begin{split}
\Psi(x,t) &\approx \left(\frac{m}{ i t}\right)^{1/2} \exp\left[\frac{i}{\hslash} \frac{(m x/t)^2}{2m} t\right] \\ 
&\times \left(\frac{\sigma^2}{\pi \hslash^2}\right)^{1/4} \exp\left[ - \frac{(m x/t)^2 \sigma^2}{2 \hslash^2}\right],
 \end{split}
\end{equation}
where one readily recognizes in the second line the momentum wave function $\tilde\Psi(m x/t)$ from \eref{2wfns}. Then, since $m x/t \equiv p$ from the classical condition imposed by the propagator one obtains, for $x=x_f$,  exactly the IT of \eref{onedIT}.

The condition for validity of the IT is that the length $(\hslash t/m)^{1/2}$ be greater than the length $\sigma$. Then let us define the beginning of the transition zone to be at the position $x_i = (\hslash t_i/m)^{1/2} = f\,\sigma$, where $f$ is a number much larger than unity. Taking the mean momentum of the wave packet components to be given by 
$\bar p = \hbar/\sigma$ gives a mean kinetic energy of $\bar E = \hbar^2/(2m\sigma^2)$. The condition that the semiclassical propagator is valid is that $\bar E\,t_i \gg \hbar$. Substituting for $t_i$ gives the condition $f \gg \sqrt 2$, which is essentially the same as the IT validity condition. Note that for fixed $\sigma$ the joint condition is independent of the mass of the particle.

As realistic examples consider the dissociation of the $H_2$ molecule into two $H$ atoms or the ionization of an electron from the ground state of the hydrogen atom. In both cases  the spatial wave packet produced will have an initial width of $\sigma \sim 1$ Bohr radius, or 1 atomic unit (a.u.) of length. If one takes the large value of $f = 100$, then the transition zone for validity of the IT begins already at the microscopic distance $x_i \sim 100$ a.u. The corresponding time for the electron to reach $x_i$ is $t_i = 10^4$ a.u.\ and for the proton, with a mass $\sim \!  10^3$ larger is 
$\sim  \!  10^7$ a.u. However, since $1$ a.u.\ of time is $\sim \!  10^{-17}$ secs., these are microscopic times. 
Careful time of flight experiments should be able to map this quantum to classical transition \cite{Helm2014}.

The satisfaction of the limit for the validity of the IT already for microscopic times and distances implies that the quantum wave function assumes a form leading to a classical interpretation of the results of measurement of position and momentum and correspondingly for any observable quantity composed of them. According to \eref{Psifree}, the spatial wave function spreads as a function of $x$ as $t$ increases. However, considered as a function of $x/t$ it remains of microscopic extent. This is the effective localization occurring as a consequence of Schr\"odinger propagation beyond the transition zone. As a corollary of \eref{ITprob} one has also the result, true generally, that 
\begin{equation}
|\Psi(\vek x'_f,t'_f)|^2 \, d\vek x'_f = |\Psi(\vek x_f,t_f)|^2 \, d\vek x_f
\end{equation}
for any two points along the classical trajectory.

According to decoherence theory the transition from quantum to classical behavior is due to the generation of localized narrow wave packets, that is, the wave function itself assumes the attribute of confined spatial extent of a classical particle. Since neither the wave function nor the density matrix is directly observable, a localized form of the wave function, whilst perhaps sufficient to infer classical behavior, is not necessary  for measured properties to change according to classical mechanics, as we have shown here. In fact
the IT shows that as soon as quantum particles leave a volume of microscopic dimensions in which their  accumulated action has become much greater than $\hslash$, their probability of detection is identical to the probability of the particle being launched effectively from position $\vek x_i = 0$ with an initial classical momentum $ \vek p_i$ which ordains the particle to arrive at a macroscopic  (detection) position $\vek x_f$ with momentum $ \vek p_f$ decided by the classical trajectory connecting these phase-space points. Detection of  individual particles at different macroscopic distances and times will lead an observer to infer a particle trajectory according to classical mechanics. In short, on the macroscopic scale the quantum world appears classical. Nevertheless, if the quantum wave function develops according to the Schr\"odinger equation, then measurements involving coherent contributions from more than one path will show full quantum interference. 

Our treatment has emphasized the detection of atomic or molecular fragments, typically as emanating from some microscopic reaction volume. In such experiments one strives to keep the interaction with, for example, residual gas molecules and stray fields of the environment to a minimum. However, even in a complete vacuum the aspect of the transition quantum to classical considered here occurs spontaneously as a result of propagation to macroscopic distances. Of course, to the extent that there is residual interaction with an environment there will be corresponding additional changes to the wave function as demanded by standard decoherence theory. Most notably, the decoherence mechanism destroys the quantum coherence between the contributions of different classical trajectories to the semiclassical density matrix, as has been shown in many applications.

JMF acknowledges the ongoing support of the Department of Energy, Chemical Sciences, Geosciences and Biosciences Division of the Office of Basic Energy Sciences.


\begin{thebibliography}{}


\bibitem{decohere} Of the voluminous literature, see for example E. Joos, H. D. Zeh, C. Kiefer, D. Guilini, J. Kupsch, and I.-O. Stamatescu, \emph{Decoherence and the Appearance of a Classical World in Quantum Theory},  2nd Ed. (Springer, New York, 2003) and references therein; W. H. Zurek, Physics Today {\bf67}, 44 (2014) and Los Alamos Science, Number 27 (2002)  and references therein; 
J. J. Halliwell Phys. Rev. D {\bf 39}, 2912 (1989). 

\bibitem{semicl} Again, from the very extensive literature, see for example R. P. Feynman and A. R. Hibbs, \emph{Quantum Mechanics and Path Integrals}, (Dover, 2010); M. C. Gutzwiller, \emph{Chaos in Classical and Quantum Mechanics}, 2nd Ed. (Springer, New York, 1990); M. V. Berry and K. E. Mount, Rep. Prog. Phys. {\bf35} 315 (1972); 
W. H. Miller J. Chem. Phys. {\bf53} 1949 (1970); J-M. Rost,  Phys. Rep. {\bf297}, 271 (1998).

\bibitem{Brumer} G. Campolieti and P. Brumer, Phys. Rev. A {\bf 50}, 997 (1994); Phys. Rev. A {\bf53}, 2958 (1996). 

\bibitem{Kemble}{E. C. Kemble, \emph{Fundamental Principles of Quantum Mechanics with Elementary Applications}, (McGraw Hill, 1937). }

\bibitem{ITfree}{M. R. H. Rudge and M. J. Seaton, Proc. Roy. Soc. London, A{\bf 283}, 262 (1965); E. A. Solovev, Phys. Rev. A{\bf 42}, 1331 (1990); 
T. P. Grozdanov and E. A. Solovev, Eur. Phys. J. D{\bf 6}, 13 (1999); J. H. Macek in \emph{Dynamical Processes in Atomic and Molecular Physics}, (Bentham Science Publishers, ebook.com, 2012), G. Ogurtsov and D. Dowek, eds.;  M. Kleber, Phys. Rep. {\bf 236}, 331 (1994).} 


\bibitem{BriggsFeagin_IT} J. S. Briggs and J. M. Feagin, J. Phys. B: At. Mol. Opt. Phys. {\bf 46}, 025202 (2013),  Phys.\ Rev.\ A {\bf 90}, 052712 (2014),
 J. M. Feagin and J. S. Briggs, J. Phys. B: At. Mol. Opt. Phys. {\bf 47}, 115202 (2014).

\bibitem{Helm2014} J. Ullrich, R. Moshammer, A. Dorn, R. Doerner, L. P. H. Schmidt, and H. Schmidt-Boecking, Rep. Prog. Phys. 66, 1463 (2003);  M. Gisselbrecht, A. Huetz, M. LavollŽe, T. J. Reddish, and D. P. Seccombe, Rev. Sci. Instr. {\bf 76}, 013105 (2013); P. C. Fechner and H. Helm, Phys. Chem. Chem. Phys. {\bf 16}, 453 (2014). 

 
\bibitem{k_measure}
Although direct momentum measurement is unusual, Eqs.\ (\ref{ITwave}), (\ref{VVtheorem}) and (\ref{ITprob}) are readily inverted to give the form identical density $d \vek x_i/d \vek p_f$ of detecting the system with momentum $ \vek p_f$ given that it started near $\vek x_i $. Namely, 
$|\tilde\Psi(\vek p_f,t_f)|^2 d \vek p_f  \approx \,  |\Psi(\vek x_i, t_i)|^2 d \vek x_i.$ 


\end{thebibliography}
\end{document}